\documentclass{elsart}
\usepackage{latexsym}
\usepackage{epsfig}
\usepackage{natbib}

\def\url#1{#1} 

\begin{document}

\begin{frontmatter}

\title{Spectral analysis of red scattered sunlight at sunrise}

\author{Fr\'ed\'eric \snm Zagury\thanksref{email1}}
\address{02210 Saint R\'emy Blanzy, France}
\thanks[email1]{E-mail: fzagury@wanadoo.fr}
\and
\author{Mitsugu \snm Fujii\thanksref{email2}}
\address{ 4500 Kurosaki, Tamashima Kurashiki, Okayama 713-8126, 
Japan}
\thanks[email2]{E-mail: aikow@po.harenet.ne.jp}

\received{December 2002}

 \begin{abstract}
We analyze and fit visible spectra of a red horizon at sunrise.
The shape of the spectra consist of a blue continuum followed by a red bump.

The reddest spectra are well fitted by the product of a spectrum of 
extinguished sunlight (Rayleigh extinction + ozone absorption) and of $1/\lambda^{4}$.
The former is essentially the radiation field in the outer atmosphere, 
at the scattering volume location; 
the latter corresponds to Rayleigh scattering by the gas.

Moving to higher altitudes, a second component, corresponding to the 
spectrum of a blue sky, must be added.

The spectra we have obtained are similar to spectra of red nebulae, 
suggesting there may be other explanations than an emission process to the 
red color of some nebulae.
   \end{abstract} 
\end{frontmatter}
 \section{Introduction} \label{intro}
In a preceding paper \citep{sol1}, we have analyzed spectra of the sun  
observed through layers of the atmosphere at increasing optical depths.

This study helped define what the radiation field at any point of 
the atmosphere, in normal conditions, should be.
It was also remarked that the light received at the earth's from a direction 
different than that of the sun, should be the product of the 
radiation field in the atmosphere and $1/\lambda^{4}$.

In the present article, we will be interested in the nature of the 
glow horizon observed from earth at sunsrise, or sunset.
\section{Data} \label{data}
\begin{table*}[h]
\caption[]{Parameters for the spectra used in the article}		
       \[
    \begin{tabular}{|c|c|c|c|c|c|}
\hline
n$^{\circ\,^{(1)}}$ &  U.T.$^{(2)}$ & exposure$^{(3)}$  & altitude  & azimuth  & angular distances from the sun\\ 
&h:min:sec&sec&($^{\circ}$) &($^{\circ}$) &($^{\circ}$) \\ 
\hline
\multicolumn{6}{|c|}{November 2001}\\
\hline
1 &  21:15:39  & 24   & 45.35  &  270.28    &       52\\
2 &  21:18:25  &  12  & 30.59  &  269.35    &       39\\
3  & 21:20:44  &  10  & 20.87  &  269.28    &       31\\
4 &  21:26:15  &   1.0 &  7.15  &  270.30   &        21\\
5  & 21:28:36   &  1.2 & 10.57  &  270.55   &        23\\ 
\hline
\multicolumn{6}{|c|}{January 2002}\\
\hline
11 &  21:56:14  &   5   &  7.7   &  295.9 & 12\\
12  & 21:57:53  &   8   &   13.3  &   296.0 & 17\\
13 &  21:59:49  &  10   &   23.1  &   296.5 & 27\\
14 &  22:02:06  &  12   &   69.5  &   303.0 & 73\\
\hline
\multicolumn{6}{|c|}{June 2002}\\
\hline
b1  & 22:36:14  &   0.20  &  64.6  &   59.5  &   82.7  \\     
b2 &  22:37:52  &   0.22 &  64.3   &  60.0   &  82.7 \\
b3  & 22:39:27  &   0.18 &   64.0  &   60.6  &   82.7 \\
\hline
\end{tabular}    
    \]
\begin{list}{}{}
\item[$(1)$] This number is used in the text and in the figures for the 
corresponding spectra.
\item[$(2)$] Time of observation. Local time is: U.T.+$09^{h}$.
\item[$(3)$] Duration of exposure.
\end{list}
\label{tbl:data}
\end{table*}
The data was acquired at the Fujii Bisei Observatory (Bisei Okayama, 
Japan, http://www1.harenet.ne.jp/\~~aikow/).
The telescope is a $28\,$~cm reflector.
The slit width of the spectroscope is
$0.1\,\rm mm$ with a  dispersion of $32.8\,\rm nm/mm$. 
The spectral resolution is $1\,\rm nm$.
The wavelength range sampled by the spectrograph is $4400\,\rm\AA$ 
to $7150\,\rm\AA$.

For this observational program, morning grow spectra, and near zenith blue 
skys, were obtained 
during three observing runs, in the early mornings of  November 11, 2001,
January 11, 2002 and June 1, 2002.
November 2001 and January 2002 spectra are observations of the red 
horizon at sunrise, at different elevations.
The three spectra of June 2002 are blue sky, high altitude spectra.
The blue spectra are proportional one to the other, so that only one is used in the 
paper.
The sky was clear for the three observing runs.
Information on the observations is given in Table~\ref{tbl:data}.

A dark exposure is substracted from the raw spectra.
The spectra are then flat-fielded using the spectrum of a tungsten lamp.
We did not require an absolute calibration for the spectra since we 
were interested only by their shape. 
The spectra presented in the paper are all normalised by a spectrum of the sun,  
kindly furnished by  G. Thuillier \citep{thuillier03}.
The normalised spectra are set to 1.0 in their short wavelength part.
\section{Analysis} \label{ana}
\begin{figure*}[p]
\resizebox{!}{1.5\columnwidth}{\includegraphics{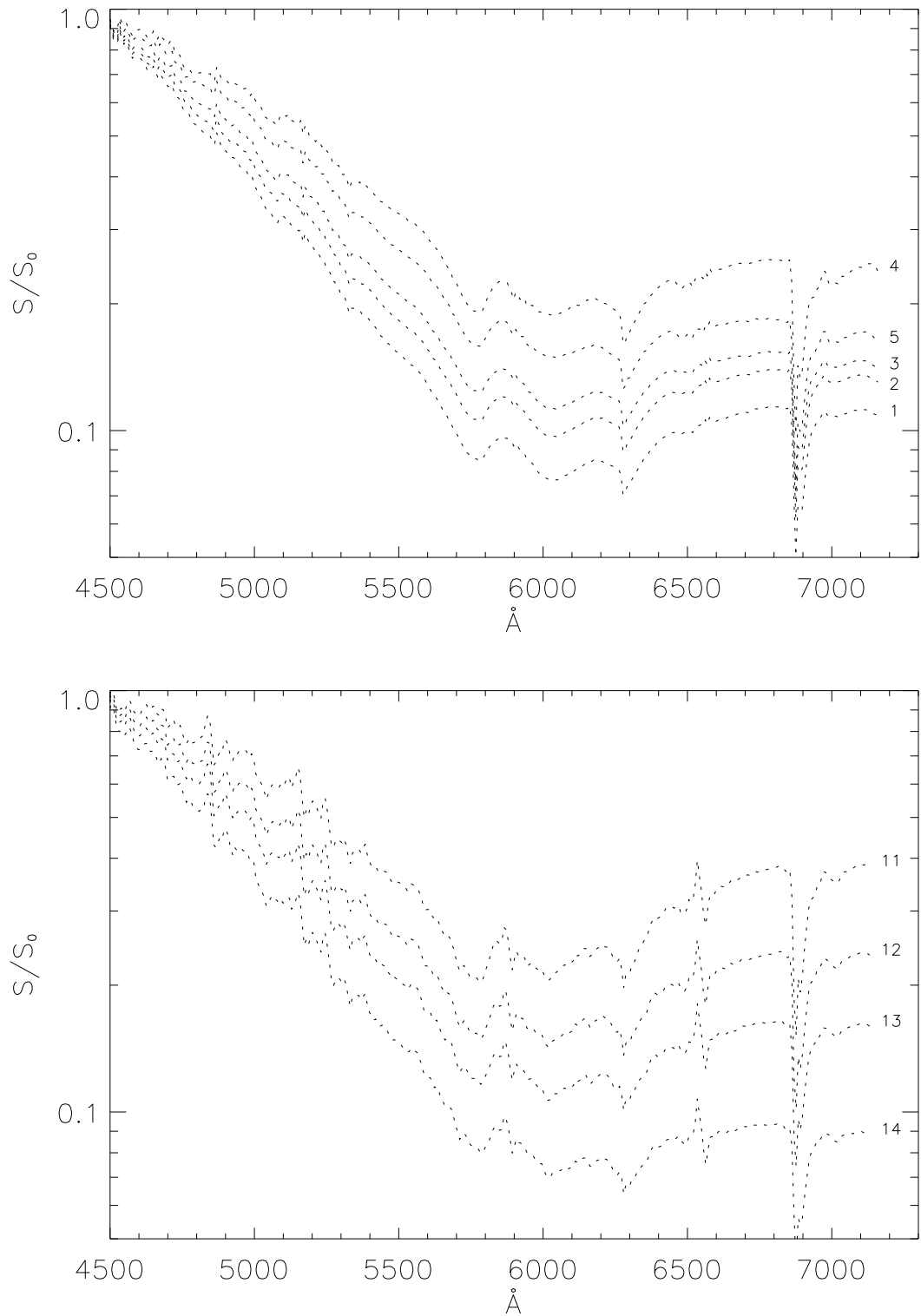}} 
\caption{Spectra of the glow at sunrise observed from Fujii Bisei Observatory.
The spectra are labeled by their number in Table~\ref{tbl:data} (first 
column). The spectra are normalized to 1.0 in the blue.
\emph{Top:} November 2001 data.
\emph{Bottom:} January 2002 data.
} 
\label{fig:rouge}
\end{figure*}
\begin{figure*}[t]
\resizebox{!}{.6\columnwidth}{\includegraphics{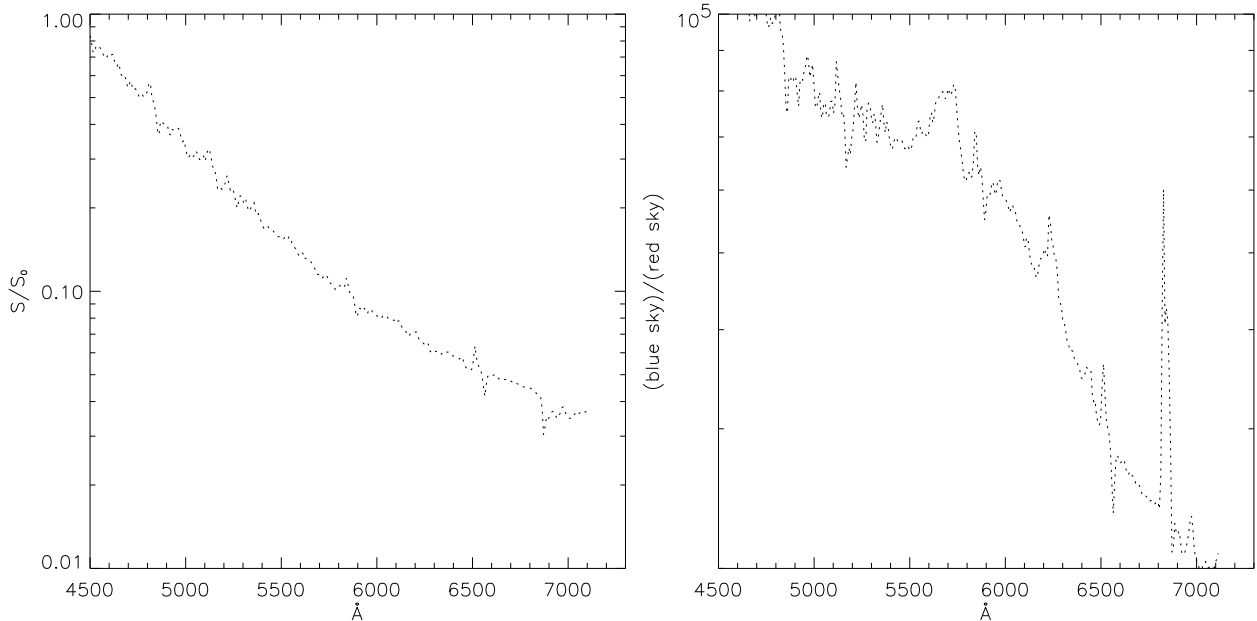}} 
\caption{Spectrum of a blue sky observed from Fujii Bisei Observatory.
\emph{left:} Spectrum~(b1) observed in June 2002.
\emph{right:} Ratio of a blue sky spectrum (spectrum (b1)) to a red sky 
spectrum (spectrum (4)).
} 
\label{fig:bleu}
\end{figure*}
\subsection{General considerations on the shape of the spectra} \label{shape}
Run~1 and run~2 spectra are separately plotted in Figure~\ref{fig:rouge}.
Moving from the near-UV to  the near-infrared, the spectra consist in 
an exponential decrease followed by a red bump.

For each run there is a straightforward relation between the altitude 
of the observation, the slope of the blue decrease,
and the importance of the red bump: the closer to the 
horizon (lowest altitude) the observation is, the smoother the 
slope, and the more important is the bump.
The impression of `red', one has when looking at the horizon at 
sunrise or sunrise, depends on the level of the bump.

The blue spectra consist of the exponential decrease solely (left plot 
of Figure~\ref{fig:bleu}); the red bump has disappeared.
These high altitude observations of a clear sky must be backscattering 
of slightly extinguished sunlight from the atmosphere along the line of sight.

For reasons of continuity, the blue part of the red spectra at the highest 
altitudes probably comprise a part of slightly 
extinguished scattered sunlight, with a spectrum similar to the blue 
light spectrum.  

The reddest spectra are the result of a longer optical paths of the sunlight 
-necessarily through the outermost part of the atmosphere- followed by scattering.
Since the outer atmosphere is mainly gas, the red spectra can be 
expected to be a mixture of gas extinction of sunlight between the sun 
and the scattering volume (Rayleigh extinction $e^{-a/\lambda^{4}}$ 
+ ozone absorption with cross-section $\sigma_{\lambda}$, $=e^{-N_{O_{3}}\sigma_{\lambda}}$), 
followed by Rayleigh scattering ($\propto 1/\lambda^{4}$).

\begin{figure*}[p]
\resizebox{!}{1.5\columnwidth}{\includegraphics{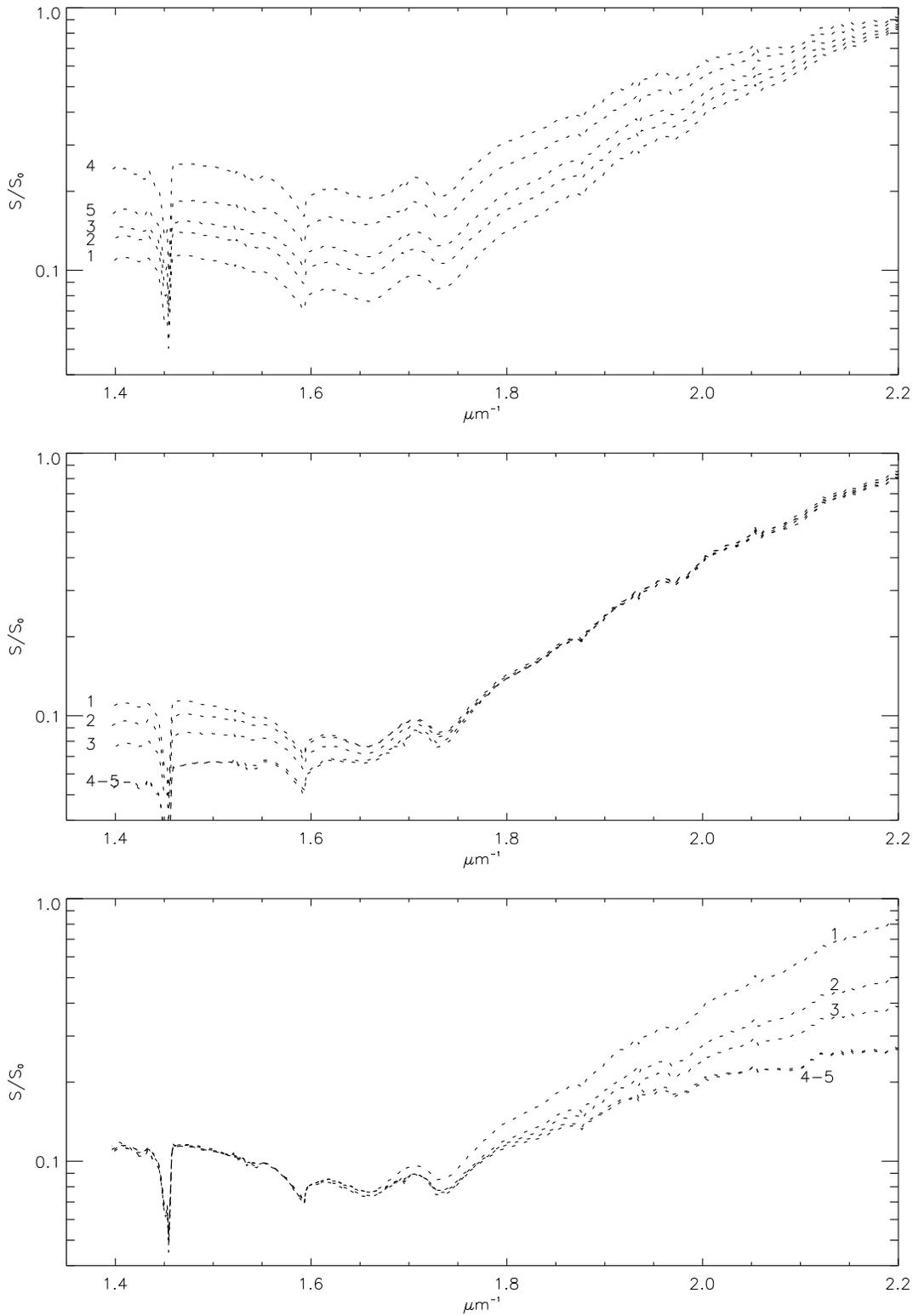}} 
\caption{
\emph{Top:} November 2001 data plotted against $1/\lambda$.
\emph{middle:} 
In the blue, spectra (2), (3), (4), (5) of November 2001 deduce one from spectrum (1) by an 
exponential of $1/\lambda$ with respective exponents 0.45, 0.7, 1.7, 1.33.
\emph{Bottom:} Same for the red part of the spectra, with exponents 
-0.45, -0.7, -0.6, -0.1.
} 
\label{fig:rel}
\end{figure*}

The presence of important absorption by ozone in the red spectra is manifest in 
the ratio of blue to red spectra (the bump in the $5000\,\rm\AA$ to $6500\,\AA$ 
wavelength region, right plot of Figure~\ref{fig:bleu}).
Empirical relations between the red spectra also confirm 
the importance of ozone absorption: the blue parts
of the red spectra deduce one from the other by an exponential 
of $1/\lambda$ (Figure~\ref{fig:rel}).
Similar relations, with exponents of opposit sign, exist between the 
red parts of the spectra.
These relations can be related to the extinction cross section of 
ozone \citep{sol1}, $\sigma_{\lambda}$, which is linear in $1/\lambda$ from $1.4\,\mu\rm m^{-1}¥$
to $1.7\,\mu\rm m^{-1}$ (with a positive slope) and from $1.7\,\mu\rm m^{-1}$ to $2\,\mu\rm 
m^{-1}$ (with a negative slope). 

It follows that the blue decreases in the red spectra can have two 
origins: blue light scattered by the inner 
parts of the atmosphere, or absorption by ozone.

Finally, the red spectra should be the sum of a red and a blue components, 
weighted by the ozone absorption.
For a given time, at a constant azimuth, 
the lowest latitude spectra should be dominated by the red light and an 
important ozone depression.
This depression and the importance of the red bump will diminish with the 
increase of altitude.
Still increasing the altitude, blue light should appear towards the near-UV 
first, and progressively replaces red light.
\subsection{Fit of the spectra} \label{fit}
\begin{figure*}[p]
\resizebox{!}{1.5\columnwidth}{\includegraphics{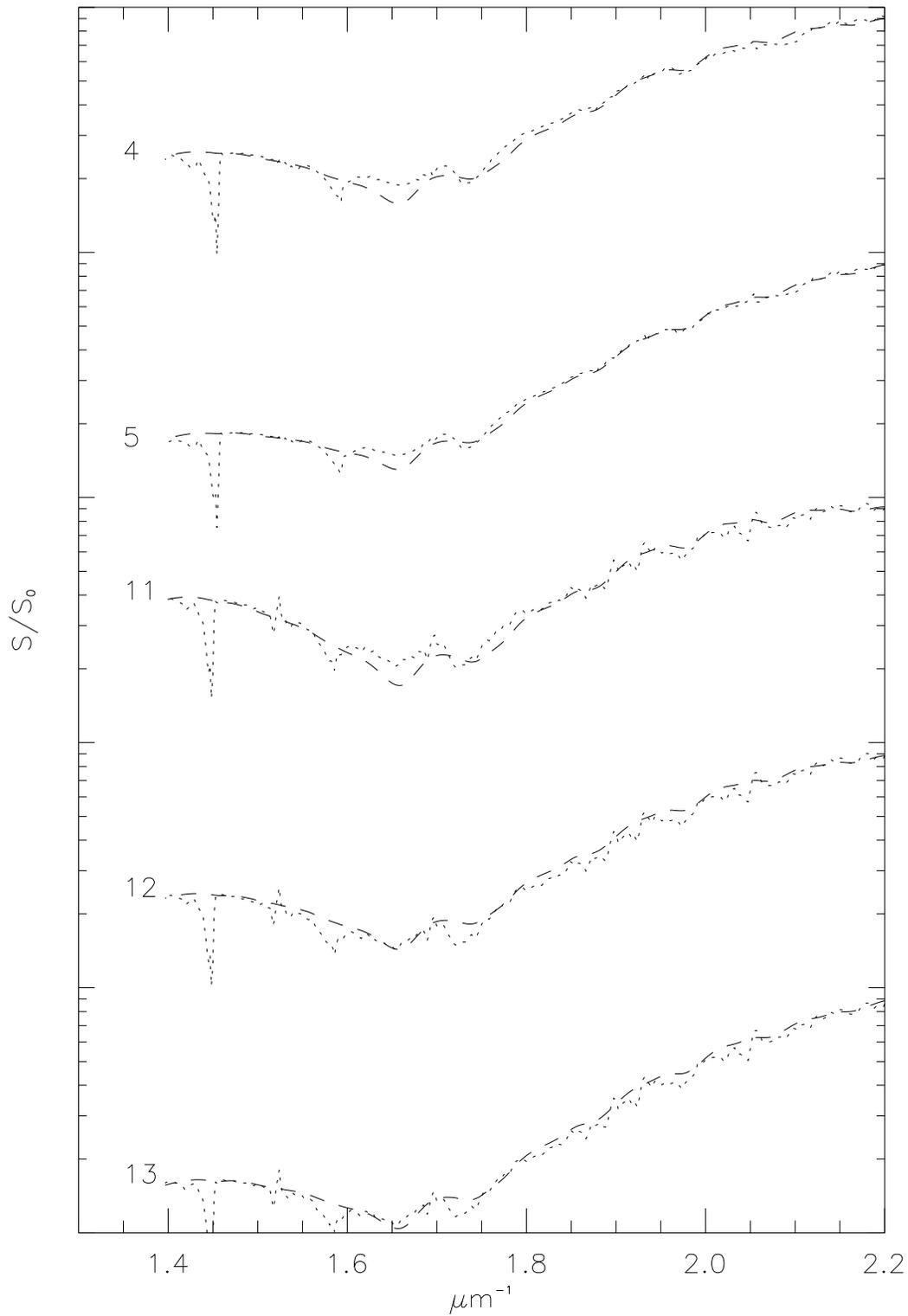}} 
\caption{
Fit of the low altitude (reddest) spectra.
The fit are a combination of Rayleigh extinction 
($\propto e^{-a/\lambda^{4}-oz(\lambda)}/\lambda^{4}$) and ozone 
absorption, in agreement with what is expected from gas extinction in 
the upper atmosphere.
} 
\label{fig:fitr}
\end{figure*}
\begin{figure*}[p]
\resizebox{!}{1.5\columnwidth}{\includegraphics{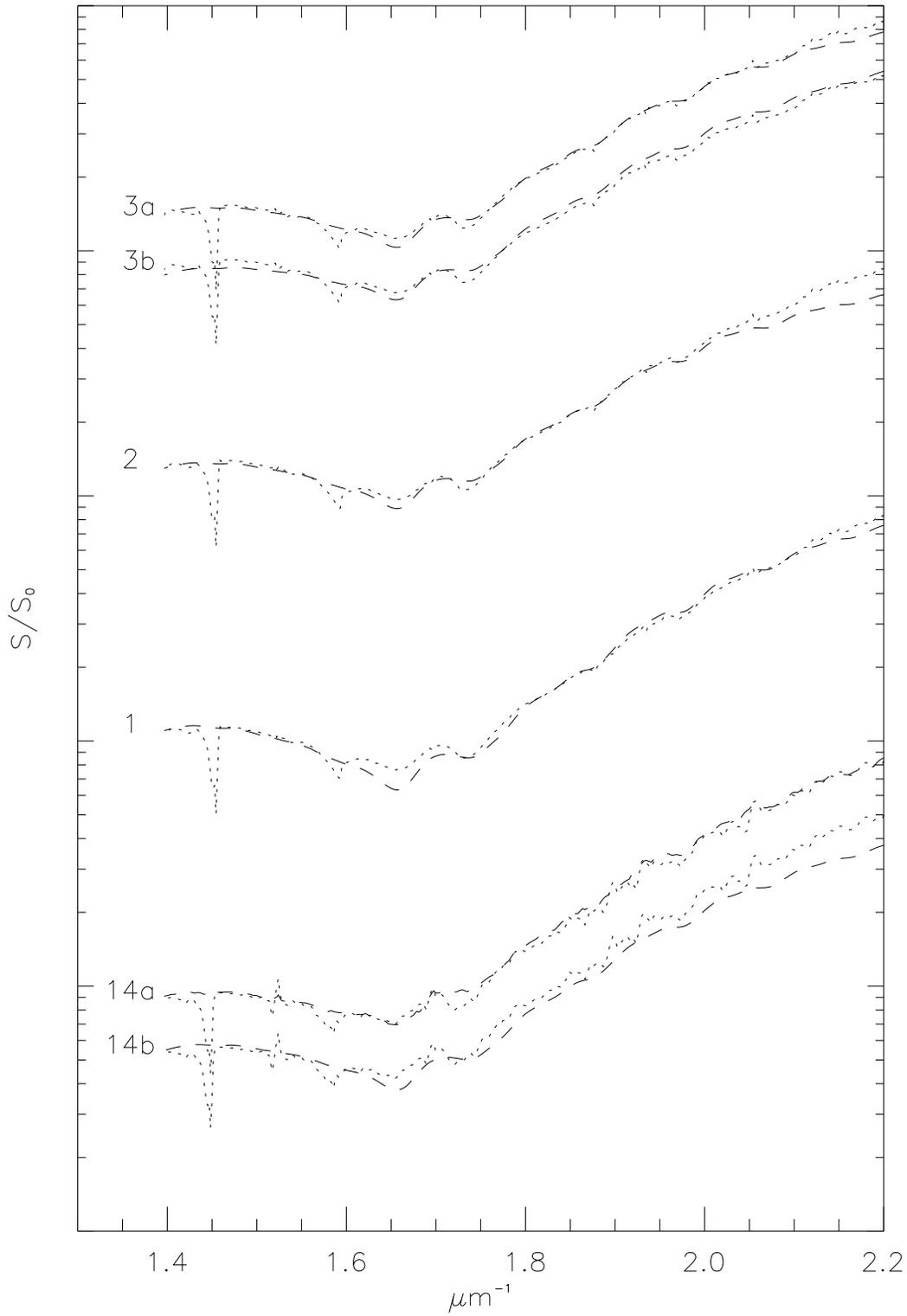}} 
\caption{
For the high altitudes spectra, the fit cannot be reduced to a one 
component gas extinction.
The fit adopted in Figure~\ref{fig:fitr} is less satisfying, as 
shown for spectrum~(3), fit~(3b).
This fit still applies to the red part of the spectra, 
but a second component is necessary to account for the blue part. 
For spectrum~(14), the fit~(14a) is a two additive components fit, 
one for the red part and an additional blue part proportional to 
blue spectrum~(b1).
} 
\label{fig:fitb}
\end{figure*}
\begin{table*}[]
 \caption[]{Fits used in Figure~\ref{fig:fitr} and Figure~\ref{fig:fitb}}		
      \[
    \begin{tabular}{|c|c|}
\hline
n$^{\circ}$ &  Fit$\,^{\,(1)}$ \\   \hline
\multicolumn{2}{|c|}{Red (lowest latitude) spectra (Figure~\ref{fig:fitr})}\\
 \hline
4 &   $\propto \,e^{-2.2\,10^{20}\sigma_{\lambda}¥}e^{-0.032/\lambda^4}/\lambda^{4}¥$ \\
5 &   $\propto \,e^{-2.0\,10^{20}\sigma_{\lambda}}e^{-0.014/\lambda^4}/\lambda^{4}¥$ \\
11 &   $\propto \,e^{-2.8\,10^{20}\sigma_{\lambda}}e^{-0.055/\lambda^4}/\lambda^{4}¥$ \\
12 &   $\propto \,e^{-2.3\,10^{20}\sigma_{\lambda}}e^{-0.030/\lambda^4}/\lambda^{4}¥$ \\
13 &   $\propto \,e^{-2.3\,10^{20}\sigma_{\lambda}}e^{-0.010/\lambda^4}/\lambda^{4}¥$ \\
\hline
\multicolumn{2}{|c|}{Red spectra with a blue sky component (Figure~\ref{fig:fitb})}\\
 \hline
1 &   $\propto \,e^{-2.7\,10^{20}\sigma_{\lambda}}e^{-0.001/\lambda^4}/\lambda^{4}¥$ \\
2 &   $\propto \,e^{-2.2\,10^{20}\sigma_{\lambda}}e^{-0.015/\lambda^4}/\lambda^{4}¥$ \\
3a &   $\propto \,e^{-2.1\,10^{20}\sigma_{\lambda}}e^{-0.011/\lambda^4}/\lambda^{4}¥$ \\
3b &   $\propto \,e^{-2.2\,10^{20}\sigma_{\lambda}}/\lambda^{4}¥$ \\
14a &   $\propto \,e^{-2.3\,10^{20}\sigma_{\lambda}}/\lambda^{4}+\alpha (b1)\,^{\,(2)}$ \\
14b &   $\propto \,e^{-2.3\,10^{20}\sigma_{\lambda}}/\lambda^{4}$ \\
\hline
\end{tabular}    
    \]
\begin{list}{}{}
\item[$(1)$] $\sigma_{\lambda}$ is the wavelength-dependent absorption 
cross-section of ozone. 
\item[$(2)$] $\alpha$ a constant. 
\end{list}
\label{tbl:recap}
\end{table*}
The reddest spectra, spectra 3, 4, 5 of November 2001 and spectra 1 and 
2 of January 2002, are well fitted by a function $\propto 
e^{-\alpha /\lambda^{4}-\beta oz(\lambda)}/\lambda^{4}$ (Figure~\ref{fig:fitr}), 
expected for gas extinction and scattering of sunlight.

It is also possible to adjust a similar fit to the higher altitude red spectra, 
but the result is not as satisfying (see spectrum~(3b) of Figure~\ref{fig:fitb}).
The best fit to the red part of these spectra
(Figure~\ref{fig:fitb}) does not account for all of the blue side.
An additional component is necessary to complete the fit, which we 
assume is the merging blue light.

Spectrum~(14a) of Figure~\ref{fig:fitb} shows the fit which can be obtained by 
adding a fraction of blue sky (blue spectrum~(1b) of June 2002) to the red fit.

We did not attempt to fit the blue spectra of June 2002, because of the 
difficulty of finding the proper analytical expression from these three blue 
spectra alone, in this limited wavelength range.
\section{Conclusion} \label{dis}
Our purpose was to understand the nature of the red light in the 
horizon at sunset, or sunrise, and to fit the corresponding spectra.

Contrary to what seems to be indicated, the shape of the spectra -a 
blue continuum followed by a red bump- 
spectra of a glow horizon  are not simply the sum of scattered sunlight by 
two media (or two different kind of particles) on the same line of sight.

The reddest (lowest atltitudes) spectra of a glow horizon are due to sunlight 
extinguished by the gas (Rayleigh extinction + ozone absorption)
in the outermost parts of the atmosphere and 
scattered -still by the gas (mainly nitrogen)- in the direction of the observer.
Extinction of the scattered light by the gas can also happen but 
should be a minor effect.
In any case, this 
will not change the analytical expression of the fit.

The radiation field at the scattering volume location is the simplest 
one observed by the  SAOZ balloon experiment \citep{sol1} and scales 
as $e^{-a/\lambda^{4}-N_{O_{3}}oz(\lambda)}$.
The scattered light received at the earth's surface is the product of this 
extinguished sunlight and of $1/\lambda^{4}$.

Towards the longer wavelengths, Rayleigh extinction is less 
important; the red spectra must vary as $1/\lambda^{4}$ (the 
$e^{-a/\lambda^{4}}$ term becomes negligible).
Thus, we can predict that the 
long wavelength part of the red bumps will vary as $1/\lambda^{4}$.

The impression of `red' has two origins.
One is due to the average slope of the spectra, more 
pronounced when moving to higher altitudes and bluer regions of the 
sky.
Second, this impression is increased by the ozone deficit feature, 
especially towards the lowest 
altitudes, because of a larger optical path and ozone absorption
in the outermost atmosphere.
The longest the optical path of sunrays at the horizon, 
the redder the spectrum will appear. 

At intermediate altitudes, in between the red light at low altitude 
and the blue near-zenith sky, we found
transition spectra which are the sum of a blue component and of a red one.

There are striking similarities between the spectra presented here and 
the spectra of some red nebulae, the best example being spectra in Orion 
observed by \citet{perrin92}.
It is currently admitted that the red color of these nebulae 
must result from an emission process.
The example of the red horizon suggests the possibility of other 
explanations for the red color of these nebulae. 
{}
\end{document}